\documentclass[apj]{emulateapj}

\usepackage{natbib}
\newcommand{\msini}{\ensuremath{M \sin{i}}}
\newcommand{\feh}{\ensuremath{[\mbox{Fe}/\mbox{H}]}}
\newcommand{\teff}{\ensuremath{T_{\mbox{\scriptsize eff}}}}
\newcommand{\persec}{\ensuremath{\mbox{s}^{-1}}}

\newcommand{\mjup}{\ensuremath{\mbox{M}_{\mbox{Jup}}}}
\newcommand{\mearth}{\ensuremath{\mbox{M}_{\mbox{Earth}}}}

\def\astrosun {\mbox{$\odot$}}
\newcommand{\Msol}{\ensuremath{\mbox{M}_{\astrosun}}}
\shorttitle{A Third Giant Planet Orbiting HIP 14810}
\shortauthors{Wright et al.}

\begin{document}

\title{A Third Giant Planet Orbiting HIP 14810}
\author{J. T. Wright} \affil{Department of
  Astronomy, 226 Space Sciences Building, Cornell University, Ithaca,
  NY 14853\\jtwright@astro.cornell.edu} 
\author{D. A. Fischer} \affil{Department of Physics and Astronomy, San
  Francisco State University, San Francisco, CA
  94132}
\author{Eric B. Ford, D. Veras, J. Wang} \affil{Department of Astronomy, University of
  Florida, 211 Bryant Space Science Center, P.O. Box 112055,
  Gainesville, FL 32611-2055}
\author{G. W. Henry} \affil{Center of
  Excellence in Information Systems, Tennessee State University, 3500
  John A. Merritt Blvd., Box 9501, Nashville, TN 37209} 
\author{G. W. Marcy, A. W. Howard\altaffilmark{1}} \affil{Department
  of Astronomy, 601 Campbell Hall, University of California, Berkeley, CA
  94720-3411}
\and
\author{John Asher Johnson\altaffilmark{2}} \affil{Institute for Astronomy, University of
  Hawai'i, Honolulu, HI 96822}

\altaffiltext{1}{Townes Fellow, Space Sciences Laboratory}
\altaffiltext{2}{NSF Postdoctoral Fellow}
\begin{abstract}
We present new precision radial velocities and a three-planet Keplerian
orbit fit for the $V=8.5,$ G5 {\sc V} star HIP 14810.  We began
observing this star at Keck Observatory as part of the N2K Planet
Search Project.  \citet{Wright07} announced the inner two planets to
this system, and subsequent observations have revealed the outer planet
planet and the proper orbital solution for the middle planet.  The
planets have minimum masses of 3.9, 1.3, and 0.6 \mjup\ and orbital
periods of 6.67, 147.7, and 952 d, respectively.  We have numerically
integrated the family of orbital solutions consistent with the data and
find that they are stable for at least $10^6$ yr.  Our photometric
search shows that the inner planet does not transit.  
\end{abstract}

\section{Introduction}

The first multiple exoplanet system detected around a normal
star\footnote{Prior to this, \citet{Wolszczan92}, detected the first
  planets outside the solar system: three extraordinary planets
  orbiting the pulsar PSR 1257+12.} was the triple system
$\upsilon$ Andromadae \citep{Butler_upsand}.  Today, over 30 systems
comprising more than one planet are known \citep{Wright09}, including
seven triple systems, two quadruple systems ($\mu$ Arae
\citep{Pepe07} and GJ 581 (Mayor et al. A\&A submitted)) and the quintuple system
55 Cancri \citep{Fischer08}.   

Individual multiplanet systems offer
insights into the dynamical evolution of planetary systems that
singleton systems cannot.  For instance, \citet{Ford05} showed that $\upsilon$ Andromedae
bears the scars of strong planet-planet scattering events preserved in
their planets' orbital parameters.  Other systems show evidence of
migration and eccentricity pumping through  mean-motion resonances
(MMRs), which may be the signpost of convergent migration in
multi-planet systems \citep[e.g.][]{KleyResonance}.  Multiplanet
systems with planet-planet interactions strong enough to be detected
at current RV precision \citep[as in GJ 876,][]{Rivera05} allow for
measurement of the inclination of the system, providing true planet
masses. 

Comparison of multiplanet systems as an ensemble to singleton systems provides
observational constraints to theories and models of the early
dynamical evolution and migration history of planetary systems.
\citet{Wright09} showed that the while the eccentricity distribution
of planets in multi-planet systems is similar to that of apparently
singleton systems, their semimajor axis distributions differ
significantly.  The concentration of planets with orbital periods near
3-days seen in the single-planet systems is absent in multi-planet
systems, as is the sharp jump beyond 1 AU in planet frequency.  The fact
that these features (the 3-day pileup and the 1 AU jump) are functions
planetary multiplicity strongly suggests that
planet-planet interactions play a key role in both migration and the
origin of eccentricities.

\section{The Tenth Triple System}

The announcement of HIP 14810 $d$ herein marks the tenth system with three
or more detected planets and only the sixth known to host three or more giant
($\msini > 10 \mearth$)
planets.\footnote{The others are $\upsilon$ Andomedae, 55 Cnc, $\mu$ Arae, HD
  69830 \citep{Lovis06}, and HD 37124 \citep{Vogt05}.}  The minimum masses and orbital periods of the planets in this
systems are similar to those of $\upsilon$ Andromedae, but with the
inner and outermost components reversed.  For the three planets of HIP
14810 we find:  $\msini =$ 3.9, 1.2, and 0.6 \mjup, and $P=$ 6.67,
147.7, and 952 d, respectively.   We find modest but significant
eccentricities for all 3 components (0.14, 0.16, and 0.17, respectively).

Table~\ref{star} contains a summary of the stellar properties of HIP
14810 (= BD+20 518), which sits at 53 pc \citep[$\pi = 18.7 \pm
  1.3$,][]{Hipparcos2} and has $V=8.5$.  We have performed an LTE
analysis of our template spectra for HIP 14810 and derived its mass
and radius using the methods described in \citet{SPOCS}.  Although HIP 
14810 is a solar mass star ($M = 0.99 \Msol$), its metallicity
($\feh = +0.26$) and evolutionary status ($\Delta M_{\rm V} = 0.63$
mag, as calculated in \citet{Wright04b}) give it a spectral type of
G5.  Its low rotation ($v \sin i \sim 0.5 \pm 0.5$ km \persec) and Ca
II H \&K activity levels ($S = 0.16$, measured with the methods described in
\citet{Wright04}) are consistent with it being an old star (age $\sim
8$ Gyr).  This combined with its relatively small 
distance from the main sequence make it a particularly good radial
velocity target, since it is expected to exhibit very low levels of
jitter \citep{Wright05}.\footnote{We expect, based on hundreds of similar stars, 2 m
  \persec\ of jitter, consistent with the residuals to our best
  fit. This value is somewhat lower than 
  predicted by the formulae in \citet{Wright05} because that work
  included instrumental sources of noise associated with the HIRES CCD
  detector in place prior to Aug 2004.  The new detector has
  significantly better charge transfer properties, and apparently
  contributes a negligible amount to our overall error budget.}

\section{Velocities and Orbital Solution}
We began observations of HIP 14810 in 2005 as part of the N2K survey 
\citep{Fischer05} at Keck Observatory using HIRES \citep{Vogt94} and
our usual iodine technique \citep{Butler96b} to achieve typical
internal (random) errors of 0.8--1.4 m\persec.   The presence of the
innermost planet and the large resuiduals to its orbital fit inspired
the California Planet Search consortium to continue regular
observations of this system at Keck. 

\citet{Wright07} announced\footnote{A preliminary orbit for the $b$
  component also appears in the {\it Catalog of Nearby Exoplanets}
  \citep{Butler06}.} the inner two planets of HIP 14810, though with a
rather poor fit for the $c$ component due to the poor phase coverage
and the unaccounted-for effects of the $d$ component.  Also
complicating the fit was what is now 
obviously a spurious data point, acquired during early dusk when
significant contamination from the Solar spectrum likely produced an
erroneous radial velocity measurement.  We have applied a more rigorous
data retention scheme (based solely on the measured internal
errors\footnote{We derive radial velocity measurements from each of many
  independent ``chunks'' of spectrum for each observation.  Our
  {\it internal error} for a given observation is determined from the
  variance of these velocities \citep{Wright05}.}, not   
deviations from a fit) to the data set presented in
Table~\ref{vels}.  These velocities and uncertainties supersede our previously published
values for this star, as we continue to refine our data reduction
pipeline \citep{Wright09}.  Note that the times given in the table are in heliocentric
Julian days, and the quoted errors are our internal (random) errors,
with no ``jitter'' included. 

\begin{figure}
\plotone{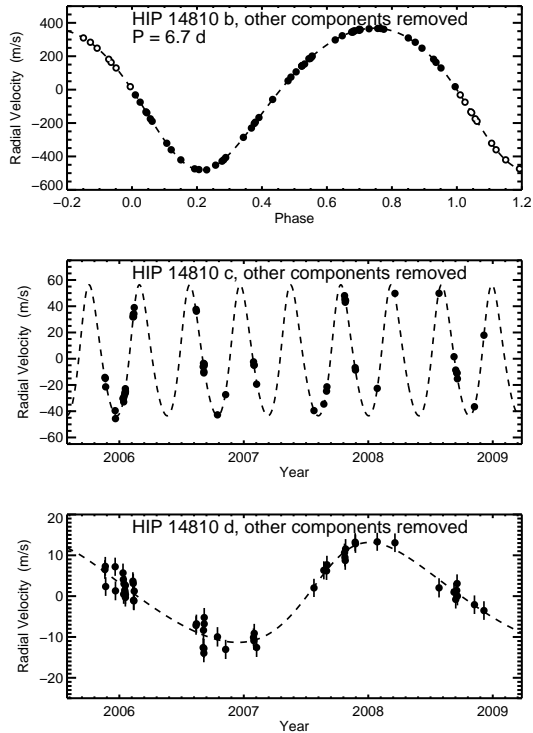}
\caption{Radial velocity curves for the HIP 14810 triple system.\label{fig}}
\end{figure}

By 2007, residuals to a 2-planet fit clearly showed coherent structure
indicative of an outer companion.  As Figure~\ref{fig} shows, by late 2008, these residuals
appeared to describe one complete orbit of a $P \sim 950$ d planet
with modest eccentricity ($e \sim 0.2$).  We have performed a Monte
Carlo false alarm probability (FAP)
analysis of the complete set of residuals to determine the likelihood that an
orbital fit of this quality could have been arrived at by chance.
This method is very similar to the FAP analyses in
\citet{Butler06,Wright07,Butler09,Howard09}, and we refer the reader to those
works for more details.  After binning the data in 24-hour intervals
and subtracting the best 2-planet fit, we redrew these 
residuals 1000 times (that is, we kept the times of observation
the same for each trial, but at each time assigned a new velocity and velocity
uncertainty pair randomly drawn from the entire set, with
replacement).  We then added the nominal orbital solution for the $c$
component back into these new residuals and performed a thorough search for the best-fit
2-planet orbital solution to each of these 1000 realizations of the
data.  We measure the FAP as the fraction of these realizations for
which we find a solution superior or equivalent in fit quality to the
nominal solution.

Note that this is an extremely conservative test of the FAP of the new
planet, $d$, because we have not
restricted the parameter search for the FAP trials to long period or
low eccentricity planets.  The best-fit solutions found for the
artificial data sets are thus often at short periods ($P<3$d) and/or high
eccentricites $e > 0.7$ \citep[see the discussion in][]{Butler09}.  Nonetheless, even with this
large parameter space available, we find that none of our 1000
realizations produced a better fit than the actual data,
yielding an FAP $<0.1\%$.

We have fit the data using the publicly available multi-planet
RV-fitting IDL package \texttt{RV\_FIT\_MP}, described in
\citet{Wright09b}.  In Table~\ref{orbital} we  
present the best 3-planet Keplerian (kinematic) fit, which yields
r.m.s. residuals of 2.3 m\persec, and we plot the fit and velocities
in Figure~\ref{fig}.  We have observed just over one
complete orbit for the outer component, HIP 14810$d$, and so its
orbital parameters are sensitive to the assumption that 
there are no additional, external planets detectably influencing the
velocities.  In particular, it is possible that such an additional
planet could be contributing a nearly linear trend to the observed radial
velocities, a trend which might be absorbed into the $d$ component's
orbital parameters as an inflated eccentricity.  Any such degeneracy will
be broken in the near future as the $d$ component completes its second orbit.
Continued observation of this system will thus reveal the presence of
any additional detectable planets and allow for further analysis of
their interactions \citep[e.g.][]{Ford05b,Wright08}.  

\begin{figure}
\plotone{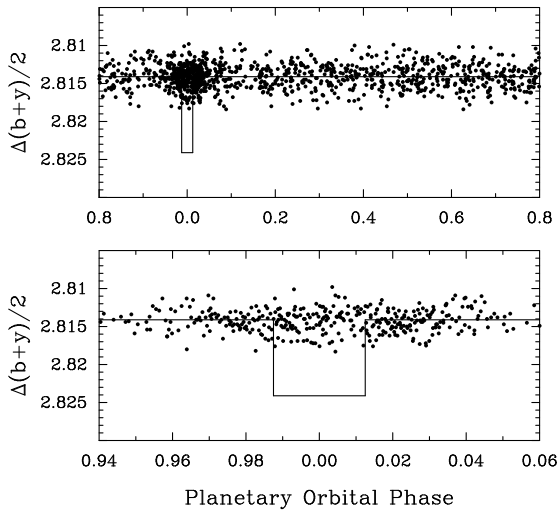}
\caption{$Top~Panel$: The 1099 photometric observations of HIP~14810 
in the combined Str\"omgren $(b+y)/2$ passband, acquired with the T11 0.8m 
APT over four observing seasons and plotted modulo the 6.673855-day orbital 
period of the inner planet HIP~14810b.  Phase 0.0 corresponds to the 
predicted time of mid transit.  A least-squares sine fit at the orbital 
period yields a semi-amplitude of only $0.00008 \pm 0.00006$ mag.  
$Bottom~Panel$:  The photometric observations of HD~14810 near the 
predicted time of transit replotted with an expanded scale on the abscissa.
The solid curve shows the depth (0.01 mag) and duration ($\pm0.0125$ phase 
units) of a central transit, computed from the orbital elements and the 
planetary and stellar properties.  The uncertainty in the predicted transit
time is smaller that the diameter of the plotted points in the lower panel.  
The brightness levels of HIP~14810 within and outside the transit window 
agree to 0.00013 mag.  Therefore, transits of HIP~14810b are ruled out for 
all reasonable densities.\label{phot}} 
\end{figure}

\section{Dynamical Modeling and Stability Analysis}

We have performed long-term numerical integrations to test for
stability of the orbital solution here, under the assumption that
there are only three planets in the system.  For these long-term
stability tests, we applied direct n-body integrations to 100 orbital
solutions consistent with the RV data, assuming an edge-on
orientation.\footnote{We have estimated the uncertainties in the orbital
  parameters in Table~\ref{orbital} using a variant of the bootstrapping described in
  \citet{Butler06}.  We generate a suite of plausible data sets and
  calculate the best-fit solution for each realization.  We selected
  100 of these realizations as the basis for this stability analysis, as
  in \citet{Wright09}.}  We
integrated for at least $10^{8}$ years using 
the hybrid integrator in {\tt Mercury} \citep{Chambers99}.  For the
majority of each integration, {\tt Mercury} uses a mixed-variable
symplectic integrator \citep{Wisdom91} with a time step approximately equal
to a hundredth of the Keplerian orbital period calculated at a
semi-major axis equal to the pericenter distance of the closest
planet.  During close encounters, {\tt Mercury} uses a Bulrich-Stoer
integrator with an accuracy parameter of $10^{-10}$.  We identified each
set of initial conditions as an unstable system if: 1) two planets
collide, 2) a planet is accreted onto the star (astrocentric distance
less than 0.005AU), or 3) a planet is ejected from the system
(astrocentric distance exceeds 100AU).  

All of our simulations proved stable, and we manually verified that in
all cases the final orbits were qualitatively similar to the initial
conditions.  We also ran an additional set of simulations for inclined
(but coplanar) orientations of the system (i.e. smaller values of
$\sin i$, and thus higher true planet masses and larger planet-planet
interactions) and find the system to be stable for all tested
scenarios with $i > 3^\circ$.  

In addition, we performed dynamical analyses of three planet solutions
using ensembles of initial conditions generated via Bayesian posterior
sampling methods \citep[e.g.][]{Ford06}, both ignoring and
including the planet-planet interactions.  We found that the
superposition of Keplerians approximation is a very good approximation
for the nominal edge-on configuration.  As the orbits approach
face-on, the planet-planet interactions eventually become significant
for inclinations of a few degrees.  While we only considered coplanar
configurations, we expect that our results are likely representative
for relative inclinations of up to 20 or 30 degrees. 

\begin{deluxetable}{cc}
\tablecolumns{2} \tablecaption{Stellar Properties of
  HIP~14810\label{star}} \tablehead{\colhead{Parameter} &
  \colhead{Value}} \startdata Spectral Type & G5 V \\
 RA & 03$^{\mbox{h}}$11$^{\mbox{m}}$14$.\!\!\!\!^{\mbox{
    s}}$230\\
 Dec. & +21$^\circ$05\arcmin 50\farcs49\\
 B-V & 0.78 \\
 V  & 8.52 \\
 Distance (pc) & 52.9 $\pm$ 4.1\\
 M$_{\mbox{V}}$ & 4.9
\\
 \teff (K) & 5485 $\pm$ 44 \\
 $\log{g}$ $[$cm$\mbox{s}^2]$ & 4.220
$\pm$ 0.06 \\
 \feh & +0.26 $\pm$ 0.03 \\
 $v\sin{i}$ & 0.54 $\pm$ 0.5
km \persec\\
 Mass (\Msol) & 0.99 $\pm$ 0.04 \\
 Radius (R$_{\astrosun}$) & 1.0 $\pm$ 0.06 \\
 S & 0.16 \\
 $\log R^\prime_{\rm HK}$ & -5.01 \\
 $\Delta M_{\rm V}$ (mag) & 0.64
\enddata
\end{deluxetable}

\begin{deluxetable}{ccc}
\tablecolumns{3} \tablecaption{Radial Velocities for
  HIP~14810\label{vels}} 
\tablehead{\colhead{HJD} & \colhead{Velocity} & \colhead{Uncertainty} \\
 \colhead{JD-2440000} &
  \colhead{m \persec} & \colhead{m \persec} }
\startdata 
13693.76627 & -147.5 & 1.1 \\
13694.83716 & -489.3 & 1.2 \\
13695.91489 & -242.3 & 1.1 \\
13723.79085 & 146.4 & 1.0 \\
13724.69303 & 304.8 & 1.2 \\
13746.81748 & -13.1 & 1.1 \\
13747.85573 & -450.0 & 0.9 \\
13748.73690 & -448.5 & 1.2 \\
13749.74186 & -86.0 & 1.1 \\
13751.90068 & 342.3 & 1.0 \\
13752.80978 & 225.5 & 0.8 \\
13753.69384 & -96.2 & 1.1 \\
13753.81261 & -154.0 & 1.0 \\
13753.90329 & -195.5 & 1.2 \\
13775.83626 & -256.0 & 0.9 \\
13776.81286 & 107.0 & 1.2 \\
13777.72301 & 332.0 & 1.2 \\
13778.72061 & 398.1 & 1.1 \\
13779.74412 & 220.8 & 1.2 \\
13961.13028 & -293.8 & 1.1 \\
13962.13341 & -426.9 & 1.0 \\
13981.97179 & -492.7 & 1.1 \\
13982.94950 & -217.1 & 1.2 \\
13983.98363 & 137.6 & 1.0 \\
13984.09915 & 169.9 & 1.2 \\
13984.98803 & 333.0 & 1.1 \\
13985.10437 & 343.6 & 1.2 \\
14023.96949 & 88.2 & 1.3 \\
14047.98188 & -396.5 & 1.2 \\
14129.79605 & -208.9 & 0.9 \\
14130.74873 & 125.2 & 0.9 \\
14131.84182 & 334.8 & 1.2 \\
14138.80623 & 336.4 & 1.2 \\
14307.13007 & 92.8 & 1.0 \\
14336.07711 & -436.4 & 1.0 \\
14344.03162 & 34.5 & 1.0 \\
14345.14559 & 306.4 & 0.9 \\
14396.82488 & -108.0 & 1.0 \\
14397.91112 & 254.0 & 1.0 \\
14398.88586 & 411.3 & 1.1 \\
14399.89293 & 364.3 & 1.4 \\
14428.00013 & -184.5 & 1.0 \\
14428.87298 & -470.7 & 1.0 \\
14492.75897 & 356.1 & 0.9 \\
14544.74216 & 260.4 & 1.1 \\
14674.09015 & 219.2 & 1.2 \\
14718.05894 & 146.8 & 1.1 \\
14723.08578 & -434.2 & 1.0 \\
14725.92426 & 348.5 & 1.2 \\
14727.05509 & 271.3 & 1.0 \\
14727.97892 & -45.6 & 1.1 \\
14777.99836 & 67.5 & 1.1 \\
14805.83497 & 358.4 & 1.1 \\
\enddata
\end{deluxetable}

\begin{deluxetable}{cccc}
\tablecolumns{4}\tablecaption{Orbital Elements for Exoplanets in the
  HIP~14810 System\label{orbital}}
\tablehead{\colhead{Property} &\colhead{$b$} & \colhead{$c$} &
  \colhead{$d$}}
\startdata
Per ($d$) &  6.673855(19) & 147.730(65) & 952(15)\\
$T_0$ (JD-2440000) & 13694.5980(70) & 14672.2400(73) & 14317.1980(73)\\
e &  0.14270(94) &  0.164(12) & 0.173(37)  \\
$\omega$ ($^\circ$) &  159.32(38) & 329.0(2.5) & 286(19)\\
$K$ (m\persec)& 424.48(44) & 50.01(46) & 12.03(49)\\
$\msini$ (\mjup)& 3.88(32) & 1.28(10) & 0.570(52) \\
$a$ (AU) & 0.0692(40) & 0.545(31) & 1.89(11) \\
\hline
r.m.s. (m\persec) & \multicolumn{3}{c}{2.3}\\
$\chi^2_\nu$ & \multicolumn{3}{c}{1.01}\\
jitter (m\persec) & \multicolumn{3}{c}{2}\\
$N_{\mbox{obs}}$ & \multicolumn{3}{c}{53}
\enddata
\tablecomments{For succinctness, we express uncertainties using
  parenthetical notation, where the least significant digit of the
  uncertainty, in parentheses, and that of the quantity are to be understood
  to have the same place value.  Thus, ``$0.100(20)$'' indicates
  ``$0.100 \pm 0.020$'', ``$1.0(2.0)$'' indicates ``$1.0 \pm 2.0$'',
  and ``$1(20)$'' indicates ``$1 \pm 20$''.}  
\end{deluxetable}

\section{Photometric Observations}

We acquired 1099 good photometric observations of HIP~14810 during four 
observing seasons spanning 1180 days between 2005 November and 2009 
February with the T11 0.8m automated photometric telescope (APT) at 
Fairborn Observatory.  The T11 APT and its two-channel Str\"omgren $b$ 
and $y$ photometer are very similar to the T8 0.8m APT and precision 
photometer described in \citet{h99}.  

The measurements of HIP~14810 were made differentially with respect to the 
comparison star HD~18404 ($V=5.80$, $B-V=0.42$, F5 IV).  We combined the 
Str\"omgren $b$ and $y$ differential magnitudes into a single $(b+y)/2$
passband to improve the precision of each measurement.  \citet{h99} gives 
further details on the acquisition, reduction, and calibration of the 
APT data.

The standard deviation of a single observation from the mean of the entire 
dataset is 0.00158 mag, which closely matches the typical measurement 
precision with this APT.  Periodogram analysis of the full dataset finds 
no significant periodicities between one and several hundred days.  In 
particular, least-squares sine fits to the 6.673855- and 147.73-day orbital 
periods of the inner ($b$) and middle ($c$) planets yield semi-amplitudes of 
only $0.00008~\pm~0.00006$ and $0.00021~\pm~0.00006$ mag, respectively, 
confirming that stellar activity is not the cause of the radial-velocity
variations at these two periods.  Although we do not expect radial velocity
variations due to long period variations in stellar activity in such
old stars \citep{Wright08}, we have also searched for photometric
variations at the period of the outer ($d$) planet ($P=952$ d).  While
our dataset is not yet long enough to make this determination
definitively, we see no suggestion of photometric variations at this
period so far.

In the top panel of Figure~\ref{phot}, we plot the APT brightness measurements of
HIP~14810 against phases computed from the 6.673855-day orbital period of 
the inner planet and a time of mid transit, JD~2,453,693.5856~$\pm$~0.0022, 
predicted from the orbital elements in Table~\ref{orbital}.  The observations near the 
predicted time of transit are replotted with an expanded horizontal scale 
in the bottom panel of Figure~\ref{phot}.  The solid curve shows the depth (0.01 mag) 
and duration ($\pm0.0125$ phase units) of a central transit of planet b, 
computed from the orbital elements and the planetary and stellar properties.
The precisely determined orbit of HIP~14810$b$ (Figure~\ref{fig}, top panel) translates
into an uncertainty in the mid-transit time that is smaller than the plotted
points in the bottom of Figure~\ref{phot}. The mean of the 120 observations within
the transit window is $2.81423\pm0.00015$ mag; the mean of the 979 observations
outside the window is $2.81410\pm0.00005$ mag.  Thus, the brightness levels
inside and outside of the predicted transit window agree to within 0.00013
mag, allowing us to rule out transits of the $b$ component for all reasonable densities.

\section{Conclusions}
HIP 14810 is orbited by at least 3 giant planets, the outermost of
which (HIP 14810 $d$) has completed just over one orbit since we began monitoring it
at Keck Observatory as part of the N2K project in 2005.  This makes it only
the sixth system known to host more than 2 giant ($\msini > 10
\mearth$) planets.  In retrospect, the previously published orbital solution for the
middle ($c$) component was hampered by a spurious data point, poor
phase coverage and the unaccounted-for effects of the $d$ 
component.  Continued monitoring will reveal the existence of any
additional outer planets.

We have performed a dynamical analysis for a suite of orbital solutions
assuming no fourth planet, and find the system to be stable at nearly
any inclination ($i < 3^\circ$).  We have performed a photometric
survey and find that the $b$ component does not transit.

\acknowledgments

The work herein is based on observations obtained at the W. M. Keck
Observatory, which is operated jointly by the University of California
and the California Institute of Technology.  The Keck Observatory was
made possible by the generous financial support of the W.M. Keck
Foundation.  We wish to recognize and acknowledge the very significant
cultural role and reverence that the summit of Mauna Kea has always
had within the indigenous Hawaiian community.  We are most fortunate
to have the opportunity to conduct observations from this mountain.

J.T.W.\ received support from NSF grant AST-0504874.  G.W.M.\ received
support from NASA grant NNG06AH52G, and D.A.F.\ from NASA grant
NNG05G164G and the Cottrell Science Scholar Program.
G.W.H.\ acknowledges support from NASA, NSF, Tennessee State University, and
the State of Tennessee through its Centers of Excellence program.
A.W.H.\ gratefully acknowledges support from a Townes Postdoctoral
Fellowship at the UC Berkeley Space Sciences Laboratory.  E.B.F.\
acknowledges the support of NASA RSA 1326409, and E.B.F.\ and
D.V.\ acknowledge support of NASA grant NNX09AB35G.  J.A.J.\ is supported by
NSF grant AST-0702821.

We thank the anonymous referee for a helpful review.  We acknowledge
the University of Florida High-Performance 
Computing Center for providing computational resources and support
that have contributed to the results reported within this paper.  This
research has made use of NASA's Astrophysics Data System.  This
research has made use of the SIMBAD database, operated at CDS,
Strasbourg, France

\facility{{\it Facility} Keck:I}

\end{document}